\documentclass[
    ,final            
  ]
  {aipproc}

\layoutstyle{6x9}

\begin{document}

\title{$\gamma$-ray emission from Narrow-Line Seyfert 1 Galaxies and implications on the jets unification}


\author{Luigi Foschini}{
  address={INAF -- Osservatorio Astronomico di Brera, via E. Bianchi 46, I-23807 Merate (Italy)}
}

\begin{abstract}
The recent discovery by {\it Fermi}/LAT of high-energy ($E>100$ MeV) $\gamma$ rays from Narrow-Line Seyfert 1 Galaxies (NLS1s) made evident the existence of a third class of $\gamma$-ray emitting Active Galactic Nuclei (AGN), after blazars and radio galaxies. It is now possible to study a rather unexplored range of low masses ($10^{6-8}M_{\odot}$) and high accretion rates (up to the Eddington limit) of AGN with relativistic jets. A comparison with the jet emission from Galactic compact objects shows some striking similarities, indicating that NLS1s are the low-mass counterpart of blazars as neutron stars are the low-mass jet systems analogue of stellar mass black holes. 
\end{abstract}

\maketitle


Highly collimated bipolar outflows have been observed in a wide variety of cosmic sources: from the supersonic jets of protostellar and protoplanetary systems, to the relativistic jets of Galactic binaries, AGN, and Gamma-Ray Bursts (see \cite{HEPRO3,IAU275} for recent reviews). In 1997, Mario Livio proposed the conjecture that the formation of a jet should be the same in all the different classes of objects, while the emission mechanisms are dependent on the type of source \cite{LIVIO}. However, the attempts to unify jets at all scales did not result yet in anything that caught a general consensus. This ``\emph{Universal Engine}'' has still to be found.

Recently, the advent of the \emph{Fermi Gamma-ray Space Telescope} (shortly \emph{Fermi}) made it possible to add one more class of AGN to the group of sources with relativistic jets. The discovery of high-energy $\gamma$ rays from NLS1s confirmed the hints collected in previous years through radio and X-ray observations (see \cite{FOSCHINI1} for a recent review). The main peculiarities of NLS1s are a relatively low mass of the central black hole ($10^{6-8}M_{\odot}$), a high accretion luminosity, and a spiral galaxy as host. In the framework of jet studies, NLS1s are important because they could be the low-mass highly-efficient part of the jet systems in AGN, as neutron stars are for Galactic binaries \cite{FOSCHINI2,FOSCHINI3}. 

\begin{figure}[!ht]
  \includegraphics[angle=270,scale=0.37]{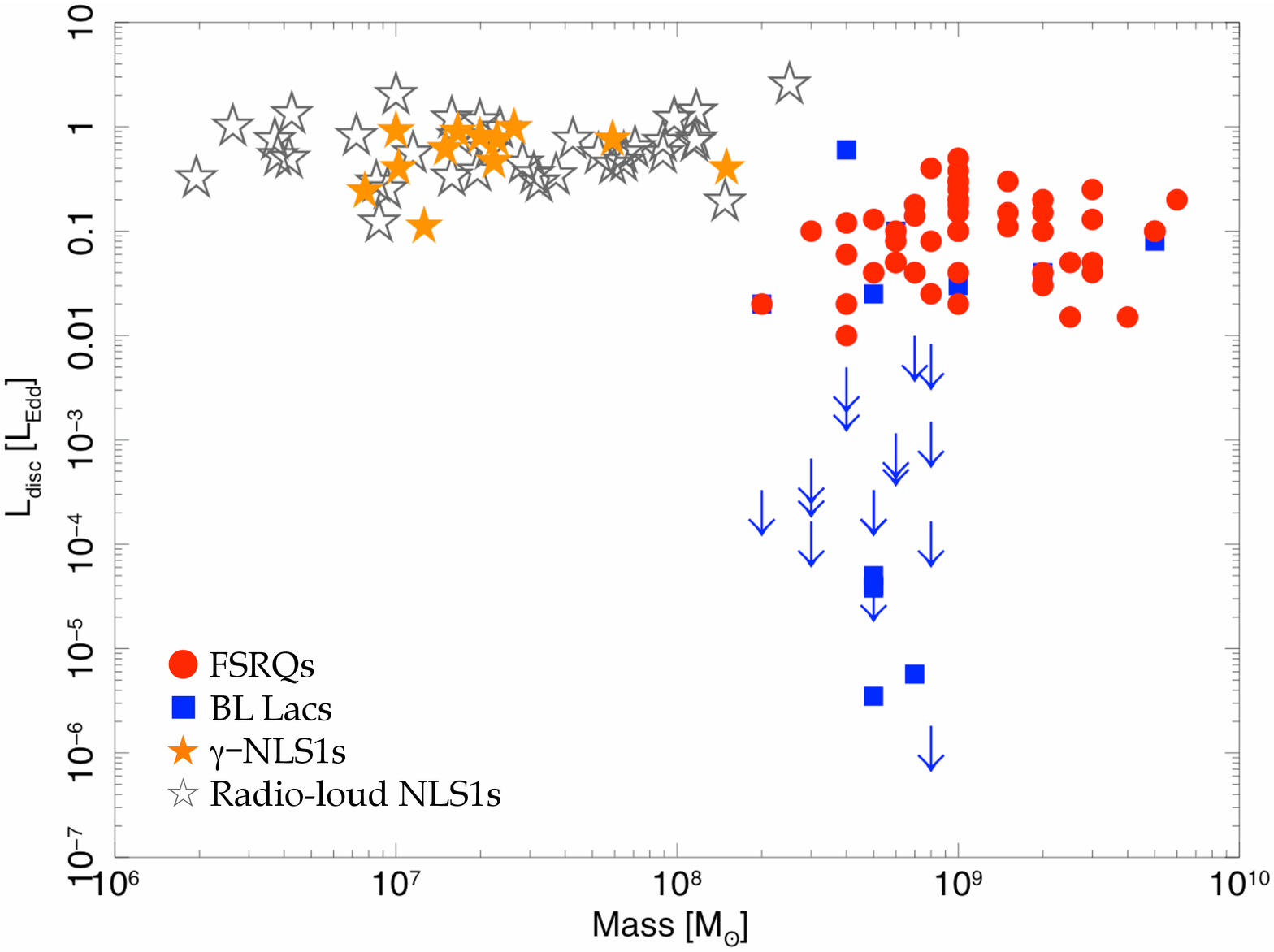}
  \caption{Disc luminosity in Eddington units as a function of the mass of the central black hole for a sample of blazars (FSRQs, red filled circles; BL Lacs, blue filled squares and arrows for upper limits of the disc luminosity), and radio-loud NLS1s (empty stars). NLS1s detected at $\gamma$ rays are displayed with filled stars.}
  \label{fig:massaccr}
\end{figure}


In this work, I outline an update of my researches in this field. The samples of sources in \cite{FOSCHINI2} were based on 39 blazars (9 BL Lac Objects and 30 flat-spectrum radio quasars, FSRQs) from \cite{GHISELLINI} with radio data from the MOJAVE Project\footnote{{\tt https://www.physics.purdue.edu/astro/mojave/}}, plus 7 NLS1s from \cite{FOSCHINI1}. In \cite{FOSCHINI3}, I have added 5 Galactic binaries (3 stellar mass black holes from \cite{corbel1,corbel2,coriat} and 2 neutron stars from \cite{migliari,miller}) in different states (80 points). Now, I have included all the blazars from \cite{GHISELLINI} (53 FSRQs, 31 BL Lacs). When high-resolution radio observation from the MOJAVE Project are not available, radio data at the highest frequency available were taken from NED\footnote{{\tt http://ned.ipac.caltech.edu/}} and converted to 15~GHz by using a flat radio spectral index ($\alpha_{\rm r}=0$). 

The sample of radio-loud NLS1s candidate to be $\gamma$-ray emitters is now made of 50 sources\footnote{See the list of sources and details of the \emph{Fermi}/LAT analysis at {\tt http://tinyurl.com/gnls1s}}. The latest quick-look analysis of about 46 months of \emph{Fermi}/LAT data resulted in 4 high-confidence detections ($TS>25$) and 8 sources detected with $9<TS<25$. 

The distribution of masses of the central black hole and disc luminosity in Eddington units for all the AGN is shown in Fig.~\ref{fig:massaccr}. All the 50 radio-loud NLS1s are shown, in order to emphasise the novelty, although only the 12 NLS1s detected at $\gamma$ rays are used for the following graphs. 

I have also included in this work one reliable candidate to be an intermediate-mass black hole (IMBH) with jet: it is HLX-1 in the galaxy ESO~243-049 ($z=0.0022$), which has an estimated mass of the compact object between $9\times 10^{3}$ and $9\times 10^{4}M_{\odot}$ \cite{WEBB} (not shown in Fig.~\ref{fig:massaccr}). 

\begin{figure}[!ht]
  \includegraphics[angle=270,scale=0.4]{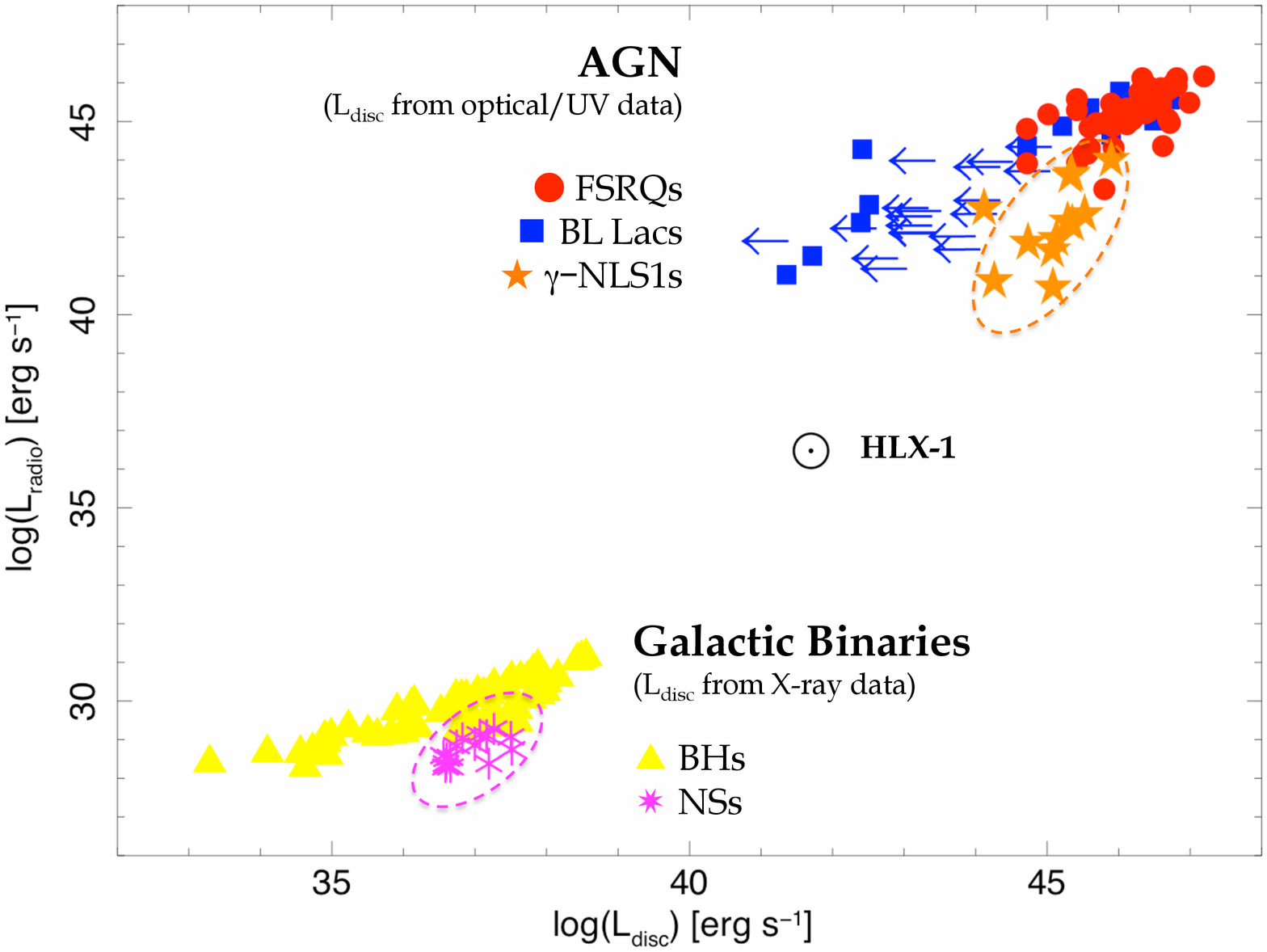}
  \caption{Jet emission at radio frequencies as a function of the disc luminosity for a sample of AGN, Galactic Binaries, and one IMBH (HLX-1, \cite{WEBB}).}
  \label{fig:jetdisc}
\end{figure}

\begin{figure}[!h]
  \includegraphics[angle=270,scale=0.4]{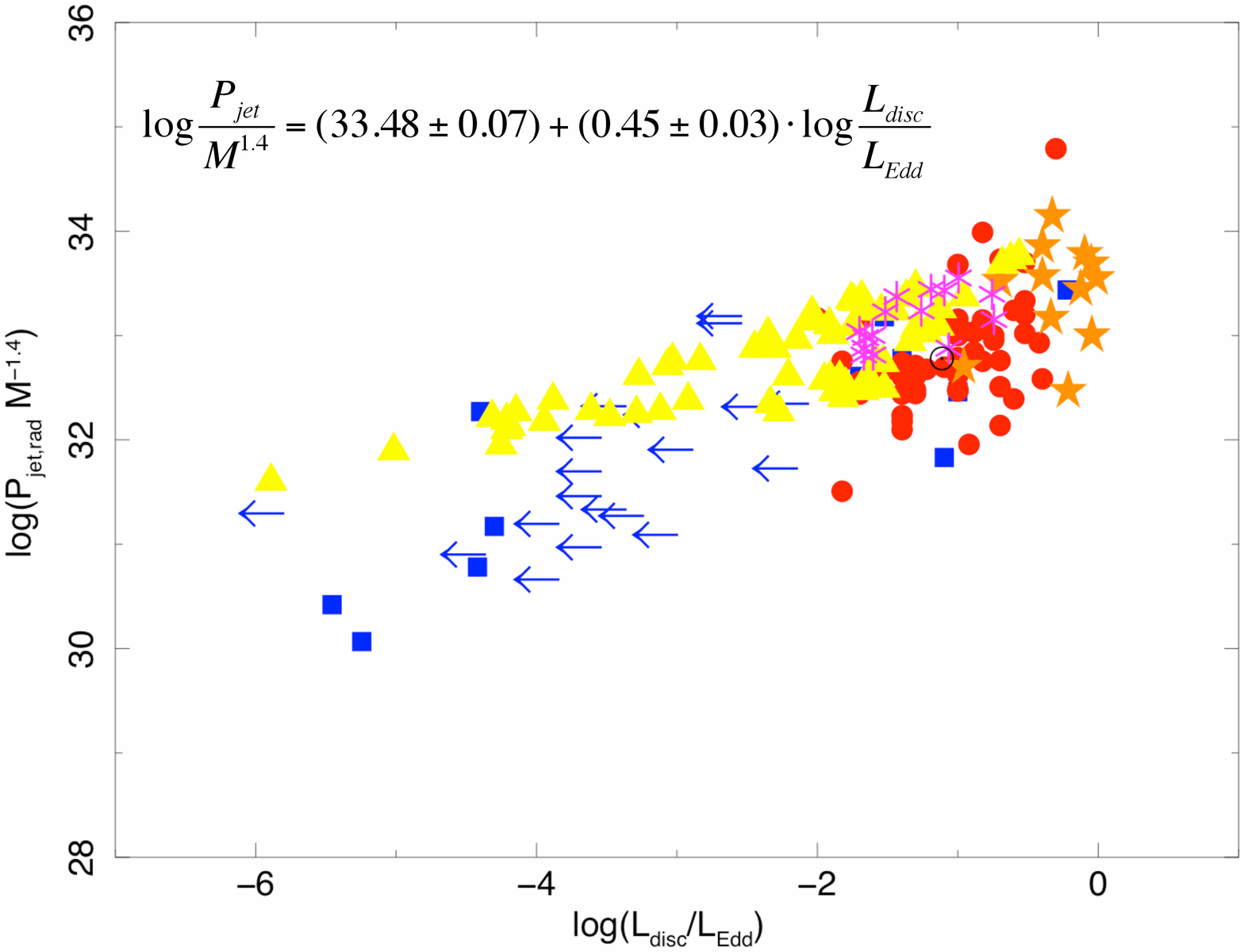}
  \caption{Jet radiative power normalised for the mass of the central compact object as a function of the disc luminosity in Eddington units for the same samples of Fig.~\ref{fig:jetdisc}.}
  \label{fig:unification}
\end{figure}


An early look at the populations under study is done by using only \emph{observed} quantities (Fig.~\ref{fig:jetdisc}). The jet emission is measured at radio frequencies (core), while the accretion disc luminosity is measured from X-rays for Galactic binaries and from optical/ultraviolet wavelengths for AGN. No model assumptions have been done, except for the standard accretion disc theory for which the emission peak depends on the mass of the compact object (X-rays for Galactic binaries; ultraviolet frequencies for AGN). In the case of the accretion power, using the same frequency range for both AGN and Galactic binaries would lead to inconsistencies given the large difference of masses. 

Fig.~\ref{fig:jetdisc} immediately shows that NLS1s are filling a region of the graph that is necessary to have a distribution of AGN similar to that of Galactic binaries. The blazars branch can overlap only with the one of stellar mass black holes, but without NLS1s there would be no AGN counterpart to the neutron stars branch. It is worth noting that HLX-1 is placed in the middle of the two distributions, as somehow expected. 

To remove the dependence on the mass of the central compact object, it is necessary to scale the disc luminosity linearly with the mass through the Eddington luminosity. Instead, the jet power has to be properly calculated before applying the scaling of $M^{1.4}$ as proved by \cite{heinz}. In the case of the first 4 discovered $\gamma$-NLS1s, the radiative jet power is taken from \cite{LAT}, which in turn is estimated by modelling the spectral energy distribution (SED) of the sources. In the other cases, I have adopted an improved version of the Eq.~(2) in \cite{FOSCHINI2}:

\begin{equation}
\log P_{\rm jet,radiative} = (12\pm2) + (0.75\pm 0.04)\log L_{\rm core,radio}
\label{eq:jets}
\end{equation}

After having scaled for the mass, all the sources are now merged into one single region, as shown in Fig.~\ref{fig:unification}. It still remains a slight dependence on the accretion power that has to be understood (cf \cite{FOSCHINI3}). It is also necessary to increase the sample of Galactic binaries, now made of just 5 sources, although sampled during several states.

\begin{theacknowledgments}
I would like to thank S. Farrell for having drawn my attention to the peculiar properties of HLX-1.
\end{theacknowledgments}

\end{document}